\documentclass[]{raa}

\usepackage{graphicx,times}
\usepackage{natbib}
\usepackage{amssymb,amsmath}
\bibpunct{(}{)}{;}{a}{}{,}

\usepackage[pagebackref=true]{hyperref}

\begin{document}

   \title{System noise temperature measurements of a 19-element phased array feed}

 \volnopage{ {\bf 20XX} Vol.\ {\bf X} No. {\bf XX}, 000--000}
   \setcounter{page}{1}

   \author{Shenghua Yu\inst{1,2},
      Dejian Ding\inst{1,8},
      Hongwei Xi\inst{1,2},
      Chengjin Jin\inst{1,2},
      Lei Yu\inst{1,2},
      Bo Peng\inst{1,3},
      Zhengwu Wang\inst{1},
      Yan Zhu\inst{1,2},
      Jianbin Li\inst{1,2,8},
      Lijia Liu\inst{1,2},
      Xinyu Zhu\inst{10},
      Junjie Zhu\inst{9},     
      Jun Wang\inst{1,2},
      Xiang Zhang\inst{1,2},
      Chao Hu\inst{1,2},
      Long Jiang\inst{1,2},
      Xingyu Chen\inst{3},
      Yaowei Li\inst{3},
      Wenjie Wang\inst{3},
      Zunli Yuan\inst{3},
      Jiangming Chen\inst{4},
      Hao Li\inst{4},
      Zhaosheng Li\inst{5},
      Xuetong Liu\inst{4},
      Chengye Li\inst{6}, 
      Zhiyong Wang\inst{6}, 
      Dangen Xie\inst{6}, 
      Hao Shang\inst{7}
   }

   \institute{ National Astronomical Observatories, Chinese Academy of Sciences, Beijing 100101, China,
   \\{\it shenghuayu@bao.ac.cn, pb@bao.ac.cn}\\
        \and
             State Key Laboratory of Radio Astronomy and Technology, Beijing 100101, China\\
	\and
Department of Physics, School of Physics and Electronics, Hunan Normal University, Changsha, Hunan 410081, China \\
\and 
Key Laboratory of Stars and Interstellar Medium, Xiangtan University, Xiangtan 411105, Hunan, China \\
\and
School of Science, Qingdao University of Technology, Qingdao 266525, China\\
\and
Bureau of Science, Industry and Information Technology, Longshan County 416800,China\\
\and
Guitang Town, Longshan County 416817, China\\
\and
University of the Chinese Academy of Sciences, Beijing, 100049, China \\
\and
Southwest University of Science and Technology, MianYang , 621010, China\\
\and
Institute for Astrophysics, University of Cologne, Z$\ddot{u}$lpicher Stra$\beta$e 77, 50937 Cologne, Germany \\
\vs \no
   {\small Received 20XX Month Day; accepted 20XX Month Day}
}

\abstract{
We present multi-epoch system noise temperature measurements of a 19-element 
hexagonal room-temperature phased array feed (PAF) using a modified Y-factor method. 
A Python-based pipeline with iterative channel calibration and maxSNR beamforming is developed 
to do data reduction, and a covariance-based decomposition scheme is proposed to extract 
mutual coupling noise temperature \(T_\mathrm{mc}\).
Long-term observations from March to June 2026 capture temporal 
receiver noise drift mainly induced by LNA temperature variations. Within the 1.37–1.47 GHz operating 
band, the measured beam-equivalent system noise temperature of the PAF is $\sim$73–89 K.
The measured \(T_\mathrm{mc}\) baseline ranges from $\sim$2 K to $\sim$8 K, 
while under thermal loading, the resonant peak of \(T_\mathrm{mc}\) can increase to more than 20 K.
The baseline of \(T_\mathrm{mc}\) is consistent with the S-parameter measurements, 
which show the \(T_{\rm mc}\) of the full array ranges less than 10.7 K. 
The baseline fluctuation of \(T_\mathrm{mc}\) most likely arises from the superposition 
of thermal and non-thermal components of the spatial electromagnetic fields.
A detachable stainless steel shield is deployed to quantify sidelobe spillover noise,  
and comparative tests show unshielded environmental interference increases receiver noise temperature 
by over 60 K. 
The proposed measurements and data processing framework provides a practical reference for noise performance 
characterization of ground-based radio telescope PAF prototypes.
\keywords{ 
instrumentation: detectors ; methods: observational ; methods: data analysis 
}
}

   \authorrunning{ }            
   \titlerunning{}  
   \maketitle

%
\section{Introduction}           
\label{sect:intro}

Phased Array Feeds (PAFs) represent the transformative focal-plane receiver 
technology for contemporary and next-generation radio telescopes, 
enabling simultaneous electronically steerable multi-beam synthesis via digital 
beamforming and substantially accelerating sky survey speed with a wide instantaneous 
field of view \citep{veidt2005,vanArdenne2009,vanCappellen2010}. 
Radio telescopes equipped with PAFs, such as the Australian Square Kilometre Array Pathfinder 
(ASKAP), Parkes Telescope, Westerbork APERture Tile In Focus (Apertif), Green Bank Telescope (GBT), 
have been deployed for observations targeting fast radio bursts, pulsars, HI survey, magnetic fields in galaxy clusters, 
continuum sources, and large-area radio sky surveys 
\citep{Hotan2021,Bannister2017,Deng2017,Li2021,vanCappellen2022,Rottgering2011,Heywood2016,Pingel2021}. 
In addition to these observations, low-noise amplifiers and cryogenic PAF hardware have also been designed 
and manufactured to meet the ultra-low noise requirements for astronomical observations \citep{Cortes-Medellin2015,Warnick2016,Groves2017,Melis2020,Burnett2020}.

Cryogenic and room-temperature L and C band PAF prototypes with various antenna types including Vivaldi, dipoles, 'rocket' 
have been developed to achieve wideband astronomical observations over the past two decades 
\citep{Verheijen2008,vanCappellen2022,Melis2020,Veidt2009,Cortes-Medellin2015,Wu2013}. 
For instance, the beam-formed system temperature normalized by aperture efficiency of cryogenic PAF for the GBT has 
been measured as $\sim$18-28 K near 1350 MHz for the boresight beam, achieving 
comparable noise temperature performance of the 1400 MHz cryogenic single-feed receiver on the GBT \citep{Roshi2018, Pingel2021}. 
The system temperature of L band room-temperature Apertif PAF is about 70K \citep{vanCappellen2022}. 

System noise temperature acts as the core sensitivity metric determining the 
detection threshold of faint cosmic radio sources, 
and its budget consists of multiple coupled noise constituents: LNA intrinsic thermal noise, antenna/cable resistive loss, 
cosmic sky background radiation, antenna sidelobes and backlobes noise, and correlated noise induced by element mutual coupling 
\citep{Warnick2008, Landon2010}. Mutual coupling, arising from near-field electromagnetic interaction between closely spaced radiators 
and beam-dependent active impedance mismatches at LNA input ports, introduces off-diagonal correlated terms in array covariance matrices 
and creates an unavoidable noise penalty for dense PAF designs \citep{Warnick2009, Warnick2005}. 
Fundamental theoretical frameworks, established 
by Warnick \& co-workers systematically model coupling-induced receiver noise degradation, derive active impedance matching criteria, and 
contrast three mainstream amplifier matching schemes \citep{Warnick2009, Warnick2008}. 
S-parameter full-wave simulation can predict inter-port transmission coefficients 
and convert coupling strength into equivalent noise temperature contributions, yet purely numerical predictions require on-telescope 
hot/cold load calibration validation \citep{Landon2010,Wu2013,Chippendale2014}. 

We in this paper conduct a quantitative diagnosis on how the overall noise performance 
including mutual coupling contributions, of the 19-element hexagonal PAF designed in \citep{Wu2013} and recently upgraded, 
varies over time and under different ambient environmental conditions. 
Moreover, we present a modified Y-factor method to decompose the mutual coupling components 
from accumulated overall system noise. 
This paper is structured as follows. Section 2 details the PAF hardware layout, hot/cold loads and shielding configuration 
for Y-factor method, full signal chain and end-to-end data processing pipeline. 
Section 3 presents measured receiver noise, and total system noise temperature against frequency and time, 
with quantitative contributions of inter-element coupling noise and spillover. 
Section 4 makes discussion of the physical mechanisms governing time-variant noise fluctuations including coupling, 
spillover and RFI effects, and comparison between our multi-epoch measurement scheme and others. 
Section 5 summarises main experimental conclusions and outlines prospective optimisations. 

\section{Method}

\subsection{Experimental configurations}
\label{subsec:EC} 
We measured the equivalent system noise temperature (\(T_{\rm sys}\)) of the 19-element PAF receiver 
using the Y-factor method, with observations conducted on four days between March and June 2026 
(March 04, April 12, May 08, and June 10). 
This method uses power measurement data obtained with a hot load and a cold load. 
The power ratio between the two states yields the ratio of their equivalent noise temperatures, 
from which we calculate the system noise budget.

During the system noise temperature measurements, a custom-built stainless steel screen was erected around the array. 
The screen was tilted at an angle of 60° relative to the ground with a height of approximately 1 m, to provide 
partial shielding of the array from noise radiated by the ground, horizon, mountain and nearby buildings. 
When the array equipped with the screen observes the background radio sky, the backend records the voltage of 
electromagnetic waves from the radio sky as cold load, which corresponds to the cold temperature in the Y-factor method.  

For the cold-load observations, we conducted additional measurements without the steel shielding screen, so that we 
could compare the results with those obtained under the shielded condition. 

As for the hot load, we assembled and fixed sixteen 0.5m$\times$0.5m$\times$0.3m pyramidal foam microwave 
absorbers on a plywood base plate to form a hot load with dimensions of \(\sim2\,\text{m} \times 2\,\text{m}\) 
and a height of \(0.34\,\text{m}\). 
At 1.5 GHz, the absorbers exhibit a minimum reflection of $-30$ dB for normal incidence, 
according to the manufacturer’s datasheet.
During operation, the hot load just covers the array equipped with the screen. 
The tips of the microwave absorbers are $\sim$2,560 mm above
the ground and $\sim$937 mm above the dipoles of the array. 

\begin{figure}
\hspace*{-0.25cm}
\centering
\includegraphics[width=16cm,clip,angle=0]{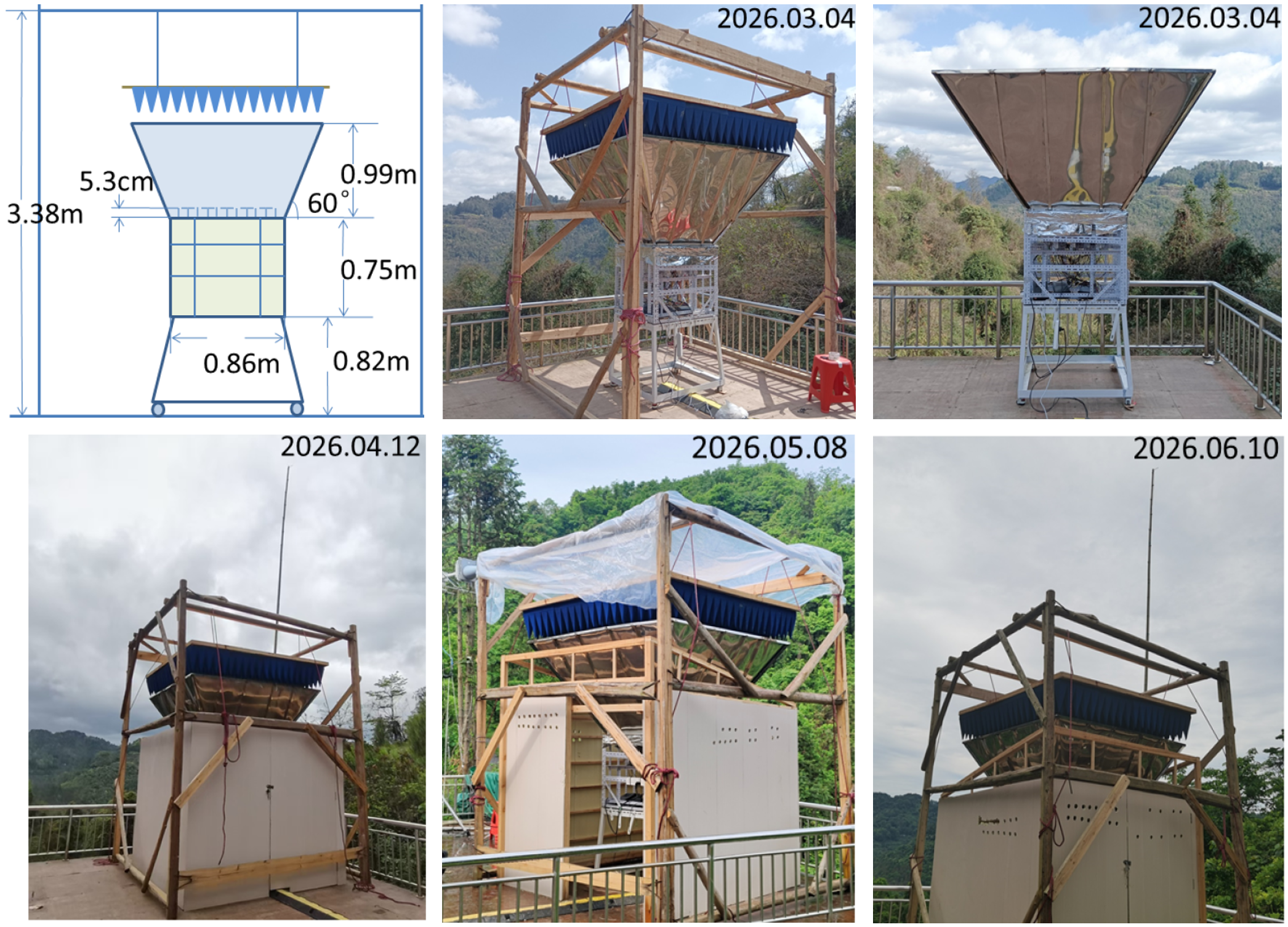}
\caption{A sketch in upper left of this figure and five photos show the sizes of the 
PAF system, the steel shielding screen, the hot-load etc., and the configurations 
during the measurements of the four days. The date is noted on the upper right of 
each photo. }
\label{fig_1}
\end{figure}

The photos and a sketch (upper left) in Figure \ref{fig_1} show the sizes of the 
PAF system, the steel shielding screen, the hot-load etc., and the configurations 
during the measurements of the four days.

\subsection{Feed Array and signal chain}
\label{sect:fa}

The PAF under test consists of 19 hexagonal arranged back-cavity dipole elements with element spacing
0.65$\lambda$($\sim$140mm), and the element number is from 0 to 18. The PAF bandwidth is 1.05-1.45 GHz with 
dual polarization \citep{Wu2013}.
The maximum radial distance from the centre of the array is 350mm. 
The 19-element array is deployed within a radio quiet zone $-$ the mountain saddle of the YinXiao Cave 
depression in LongShan, HuNan province. 

The radio-frequency signal from each PAF antenna element first enters 
a low-noise amplifier (LNA, ZX60-P162LN+). The amplified signal passes 
sequentially through a bandpass filter (BPF, 1370–1470 MHz), a second 
amplifier (ZRL-2150+), and another bandpass filter. 
After further amplification by the ZFL-2500VHX+ amplifier, the signal is 
transmitted via optical fiber (OF) using an optical transmitter module (OTM) 
and an optical receiver module (ORM).The received electrical signal goes 
through a 1370–1470 MHz bandpass filter and then enters a mixer (Mix). 
A tunable local oscillator (LO) generator provides the reference signals 
for frequency down-conversion. The intermediate-frequency output is filtered 
by a DC–40 MHz low-pass filter (LPF), then amplified by a broadband amplifier 
with 40 dB gain, 5 dB noise figure. 
A second DC–40 MHz low-pass filter suppresses high-frequency spurious components 
before the signal is fed into the digitizing data acquisition unit. 

\begin{figure}
\hspace*{-1.05cm}
\centering
\includegraphics[width=15cm,clip,angle=0]{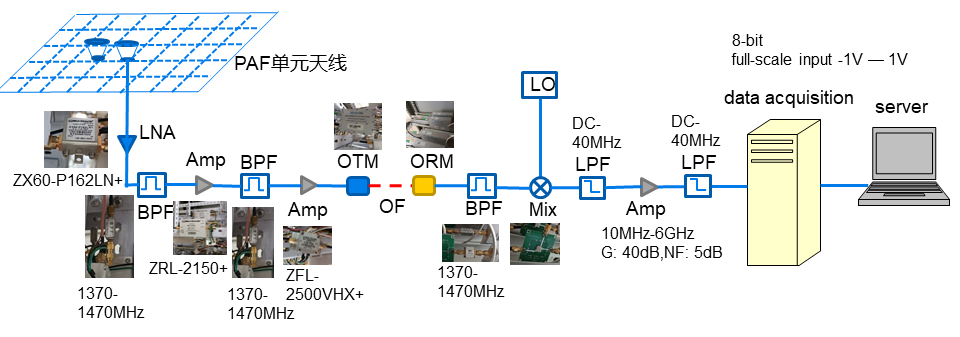}
\caption{Signal flow block diagram illustrating the electromagnetic wave receiving chain, 
from antenna input to voltage data logging on the server.}
\label{fig_2}
\end{figure}

Figure \ref{fig_2} illustrates the electromagnetic wave receiving chain, 
from antenna input to voltage data logging on the server.

\subsection{Backend and raw data preprocessing}
\label{subsec:rawdata}

We adopt the ROACH2 platform equipped with the HMCAD1511 analog-to-digital converters. 
This is an 8-bit ADC operating at a sampling rate ($f_{\rm s}$) of 77 MSps (\(f_{\rm s}=77\) MHz) 
and analog full-scale input $\pm1$V. The sampled data are transmitted to the server for offline processing.  

We record the baseband data, and calculate a spectrum using fast Fourier transform (FFT) with N=2048 sampling points and the Hanning window.  
The highest frequency of the spectrum, given by the Nyquist frequency, is $f_{\rm max}=\frac{f_{\rm s}}{2}=38.5 \rm MHz$. 
The time required to record 2048 sampling points is $t_{\rm s}=2048/(77\rm MHz)=26.5974$ microseconds. 
The minimum frequency and frequency resolution is $f_{\rm min}=\delta f=\frac{1}{t{\rm s}}=0.0375977$ MHz, 
and the channel number is 1024. In order to reduce the data size, the 1024 initial frequency channels are rebinned 
into 64 channels via power summation of every 16 channels, and the rebinned frequency interval is $\Delta f\approx$0.60 MHz.  
We in this paper calculate the system noise temperature $T_{\rm sys}$ as a function of frequency with 
the rebinned frequency interval $\Delta f$. 
The total observation time of each measurement is about 10.5s with time resolution $t_{\rm s}=26.5974$ microseconds. 

We carried out one system noise temperature measurement together with one backup observation on each of the four observing days.
Within each observing run, we collected three datasets: hot load, cold load with the metal screen, and cold load without the screen. 
Each dataset occupies approximately 15.77 GB, summing to 47.31 GB per observing run.
To extend the frequency coverage, we varied the local oscillator (LO) frequency across different measurement sessions.
The LO configurations are as follows: 1.35 GHz and 1.41 GHz on March 04; 1.35 GHz on April 12;
1.35 GHz and 1.49 GHz on May 08; and 1.35 GHz, 1.40 GHz, and 1.45 GHz on June 10.
The total size of all raw observational data is approximately 757 GB, and the storage volume of preprocessed data reaches 1.152 TB.

\subsection{Data reduction procedures} 

We developed a python-based calibration pipeline to perform consistent phase and amplitude correction 
for the 19-element phased array antenna system. The pipeline processes segmented time-domain voltage 
datasets acquired from individual antenna channels. 
The data reduction procedures are as follows:\\ 
(1) Raw data files are systematically screened to discard invalid samples, 
such as incomplete matrices, to guarantee high-quality inputs for downstream processing;\\
(2) Before calibration, segment-wise zeroing is applied to preprocessed voltage 
amplitudes via mean subtraction to remove DC baseline offsets;\\ 
(3) To minimize the system noise temperature, iterative mutual calibration is 
implemented across the array: each antenna element is sequentially selected as 
the temporary reference to correct the complex responses of all remaining 
receiving array channels;\\
(4) Fully calibrated time-series data are forwarded to the beamforming stage, 
where the maximum signal-to-noise ratio (maxSNR) criterion is utilized to solve 
for the optimal complex amplitude weighting coefficients of every antenna element.

We adopt a narrowband complex baseband signal model analogous to the formalism presented in \citet{Boonstra2003}. 
The signal bandwidth is sufficiently narrow such that inter-element propagation delays only introduce constant phase 
offsets rather than waveform distortion. 
Considering an M-element PAF targeted at a source signal $s(\rm t)$ during calibration observations, 
\(x_i(t)\) denote the narrow band complex baseband signal at frequency $f$ from the $i$-th antenna channel at time $t$.  
The single-channel signal model can be expressed as \(x_i(t) = \gamma_i e^{j \phi_i} a_i s(t) + n_i(t)\), 
where \(\gamma_i\) represents the lumped amplitude gain of the $i$-th receiver channel, accounting for cable resistive losses, 
amplifier gain variations and environmental attenuation; \(\phi_i\) denotes the cumulative instrumental phase offset originating 
from hardware delay mismatches; \(a_i = e^{-j 2\pi f \tau_i}\) represents the geometric phase shift induced by the propagation 
time delay \(\tau_i\) of the source wavefront relative to the reference antenna element; 
and \(n_i(t)\) is the uncorrelated system noise contribution from passive loss noise, LNA noise 
and thermal background spillover.

The signals from all M elements can form a column vector \(\mathbf{x}(t) = \left[x_1(t), x_2(t), \dots, x_p(t)\right]^T\), which 
equals to 
\begin{equation}
\mathbf{x}(t)=\mathbf{G}\mathbf{s}(t)+\mathbf{n}(t), 
\end{equation}
where \(\mathbf{G} = \left[\gamma_1 e^{j \phi_1} a_1,\ \gamma_2 e^{j \phi_2} a_2,\ \dots,\ \gamma_M e^{j \phi_M} a_M\right]^T\) 
is array gain vector combined with all channel-dependent complex response terms. 
The core goal of iterative mutual calibration is to solve for the unknown complex entries of \(\mathbf{G}\) by eliminating the 
inter-channel amplitude and phase inconsistencies. 

The calibration procedure consists of two steps: phase alignment and amplitude equalization. 
First, FFT-based cross-correlation is employed to achieve precise time-domain 
synchronization between each target antenna and the reference channel. After removing DC offsets from 
both the reference and target signals, the optimal temporal shift is determined from the correlation peak, 
and each waveform is circularly realigned to match the reference phase. 

An amplitude scaling factor is then calculated according to the variance ratio between the 
reference signal and each individual channel, which normalizes the fluctuation power of all 
antenna outputs to a consistent level. The relative variance error between each calibrated 
channel and the reference is quantitatively recorded to evaluate calibration performance.
This two-dimensional joint 
calibration effectively eliminates inter-channel phase delays and amplitude inconsistencies 
caused by instrumental differences and transmission variations across the array. 

In the maxSNR technique, we first solve the generalized eigenvalue equation 
\(\mathbf{R}_{\rm sn} \mathbf{u}_\mathrm{max} = \lambda_\mathrm{max} \mathbf{R}_\mathrm{n} \mathbf{u}_\mathrm{max}\)
where $\mathbf{R}_{\rm sn}$ and $\mathbf{R}_\mathrm{n}$ is the signal-plus-noise covariances and noise covariances 
obtained from the observations, $\mathbf{u}_\mathrm{max}$ the dominant eigenvector associated with the maximum eigenvalue 
\(\lambda_\mathrm{max}\) to the eigenvalue equation.  
We in this study consider the hot-load is the signal source, and the cold sky is the external noise source apart from the PAF receiver. 
In this step, we obtain the signal-of-interest covariance matrix 
\(\mathbf{R}_{\mathrm{s}} = \lambda_\mathrm{max} \mathbf{r}_s \mathbf{r}_s^H\), where 
\(\mathbf{r}_s=\mathbf{R}_\mathrm{n} \mathbf{u}_\mathrm{max}\) represents the array response, and 
the superscript \(()^H\) denotes the Hermitian conjugate (conjugate transpose) of a matrix. 
We then solve the equation
\(\mathbf{R}_\mathrm{s} \mathbf{w}_{\mathrm{maxSNR},s} = \lambda_\mathrm{max} \mathbf{R}_\mathrm{n} \mathbf{w}_{\mathrm{maxSNR},s}\)
to obtain the beamformer weight \(\mathbf{w}_{\mathrm{maxSNR},s}\) \citep{Landon2010}.

\subsection{Noise temperature budget}

The beam equivalent system noise temperature $T_{\rm sys}$ can be expressed as \citep{Warnick2008,Landon2010}
\begin{equation}
T_{\rm sys}=\eta_{\rm rad}(T_{\rm sky}+T_{\rm sp})+T_{\rm rec},
\end{equation}
where $\eta_{\rm rad}$ is the radiation efficiency of PAF and here we adopt $\eta_{\rm rad}=1$;
$T_{\rm sky} = T_{\rm cmb} + T_{\rm gb} + T_{\rm atm}$ the
sum of noise temperature in the main lobe caused by the cosmic microwave background (CMB) 
radiation $T_{\rm cmb}\approx$2.72548K \citep{Fixsen+2009}, galactic background $T_{\rm gb}\approx$1K, 
and atmospheric noise $T_{\rm atm}\approx$2K; 
We assume that the equivalent spillover noise temperature due to electromagnetic waves leak from 
sidelobes and backlobes into the receiving system is $T_{\rm sp}\approx$2K; 
$T_{\rm rec}$ is the beam equivalent receiver noise temperature. 
We have $T_{\rm rec}=\tilde{T}_{\rm rec}+T_{\rm loss} +T_{\rm mc}$, where 
$\tilde{T}_{\rm rec}$ is the noise temperature contribution of discrete electronic devices 
from the first LNA to the ADC, $T_{\rm loss}\approx$10K the loss noise 
including resistive losses in the antenna, cables, and connectors ahead of the LNA, 
dominated by the coaxial cable loss, $T_{\rm mc}$ the noise temperature due to the 
mutual coupling which will be dealt with in Eq.\,\ref{eq:tmc}.

Assuming that there is perfect impedance match between the element antenna
and the immediately following LNA, and given that noise temperature scales 
linearly with the power received by the PAF system, the equivalent PAF receiver 
noise temperature $T_{\rm rec}$ can be solved by
\begin{equation}
\frac{T_{\rm rec}+T_{\rm sky}}{T_{\rm rec}+T_{\rm hot}} 
=\frac{\mathbf{w}^{H}\mathbf{R}_{\rm cold}\mathbf{w}}{\mathbf{w}^{H}\mathbf{R}_{\rm hot}\mathbf{w}},
\label{eq:trec}
\end{equation}
where $\mathbf{R}_{\rm cold}$ and $\mathbf{R}_{\rm hot}$ are the known voltage noise covariance matrices 
calculated from the measurement data when the PAF pointed to the cold load and hot load, 
$\mathbf{w}$ the weighting factor matric of array dipole elements calculated from the maxSNR technique. 
We assume that the temperature $T_{\rm hot}$ of hot load equals to the isotropic ambient temperature.  
The measured ambient temperature was near 288 K on March 04, 292 K on April 12, 289 K on May 08, and 301 K on June 10. 
In order to easily compare the $T_{\rm sys}$ measured on the different dates, we set the ambient temperature  
to be 288 K unless otherwise specified. 

\subsection{Modified Y-factor algorithm for mutual coupling extraction}
\label{sec:myfactor}
Electromagnetic waves reflected from neighboring antenna elements, as well as reflected signals 
caused by impedance mismatches along the RF feed chains, can re-excite the array apertures and 
couple into the PAF receiver channels, leading to fluctuations in the system noise temperature. 
This phenomenon is referred to as the inter-element mutual coupling effect. 
The noise temperature contribution originating from such mutual coupling is incorporated into 
the equivalent receiver noise temperature during measurement.

We estimate the noise contribution induced by inter-element mutual coupling 
by computing the proportion $p_{\rm mc,\ast}$ of sum of the off-diagonal entries relative to the 
total sum of all entries within the cold-load covariance matrix \(\mathbf{R}_\mathrm{cold}\) and 
hot-load covariance matrix \(\mathbf{R}_\mathrm{hot}\). 
To the 19-element array, the mutual coupling noise here is defined as 
\begin{equation}
\frac{T_{\rm mc,\ast}}{\tilde{T}_{\rm rec}+T_{\rm mc,\ast}+T_{\rm loss}+T_{\ast}} 
=p_{\rm mc,\ast},
\label{eq:tmc}
\end{equation}
where the proportion $p_{\rm mc,\ast}$ is 
\begin{equation}
p_{\rm mc,\ast}=\frac{\sum_{ik}{\rm w}_{i}{\rm w}_{k}{\rm R}_{ik,\ast}}{\mathbf{w}^{H}\mathbf{R}_{\ast}\mathbf{w}}, ~~~~i,k=0,18,i\neq k 
\end{equation}
with ${\rm R}_{ik,\ast}$ being the matrix entry in $\mathbf{R}_{\ast}$, and subscript $\ast$ denoting cold or hot. 
Note that $T_{\rm cold}=T_{\rm sky}$.
This definition allows us to investigate how the mutual coupling noise varies with the cold and hot loads. 

We can calculate the \((\tilde{T}_{\rm rec}+T_{\rm loss})\) and \(T_{\rm mc,\ast}\) from Eqs.\,\ref{eq:trec}and \ref{eq:tmc}, and the result 
of the modified Y-method are 
\begin{equation}
(\frac{Y}{1-p_{\rm mc,cold}}-\frac{1}{1-p_{\rm mc,hot}})(\tilde{T}_{\rm rec}+T_{\rm loss})=\frac{T_{\rm hot}}{1-p_{\rm mc,hot}}-YT_{\rm cold}(1+\frac{p_{\rm mc,cold}}{1-p_{\rm mc,cold}}),
\label{eq:ctrec}
\end{equation}
and 
\begin{equation}
(1-p_{\rm mc,\ast})T_{\rm mc,\ast}=p_{\rm mc,\ast}(\tilde{T}_{\rm rec}+T_{\rm loss})+p_{\rm mc,\ast}T_{\ast}, 
\label{eq:tmcr}
\end{equation}
where we let \(Y=\frac{\mathbf{w}^{H}\mathbf{R}_{\rm hot}\mathbf{w}}{\mathbf{w}^{H}\mathbf{R}_{\rm cold}\mathbf{w}}\). 
When the mutual coupling can be neglected, i.e. \( p_{\rm mc,\ast}\rightarrow0 \), the Eq.\,\ref{eq:ctrec} becomes 
\( (Y-1)(\tilde{T}_{\rm rec}+T_{\rm loss})=T_{\rm hot}-YT_{\rm cold} \) for the Y-factor method. 

\section{Results}
\label{sect:results}

\subsection{Equivalent receiver noise}
\label{subsec:trec}
\begin{figure}
\hspace*{-1.05cm}
\centering
\includegraphics[width=18cm,clip,angle=0]{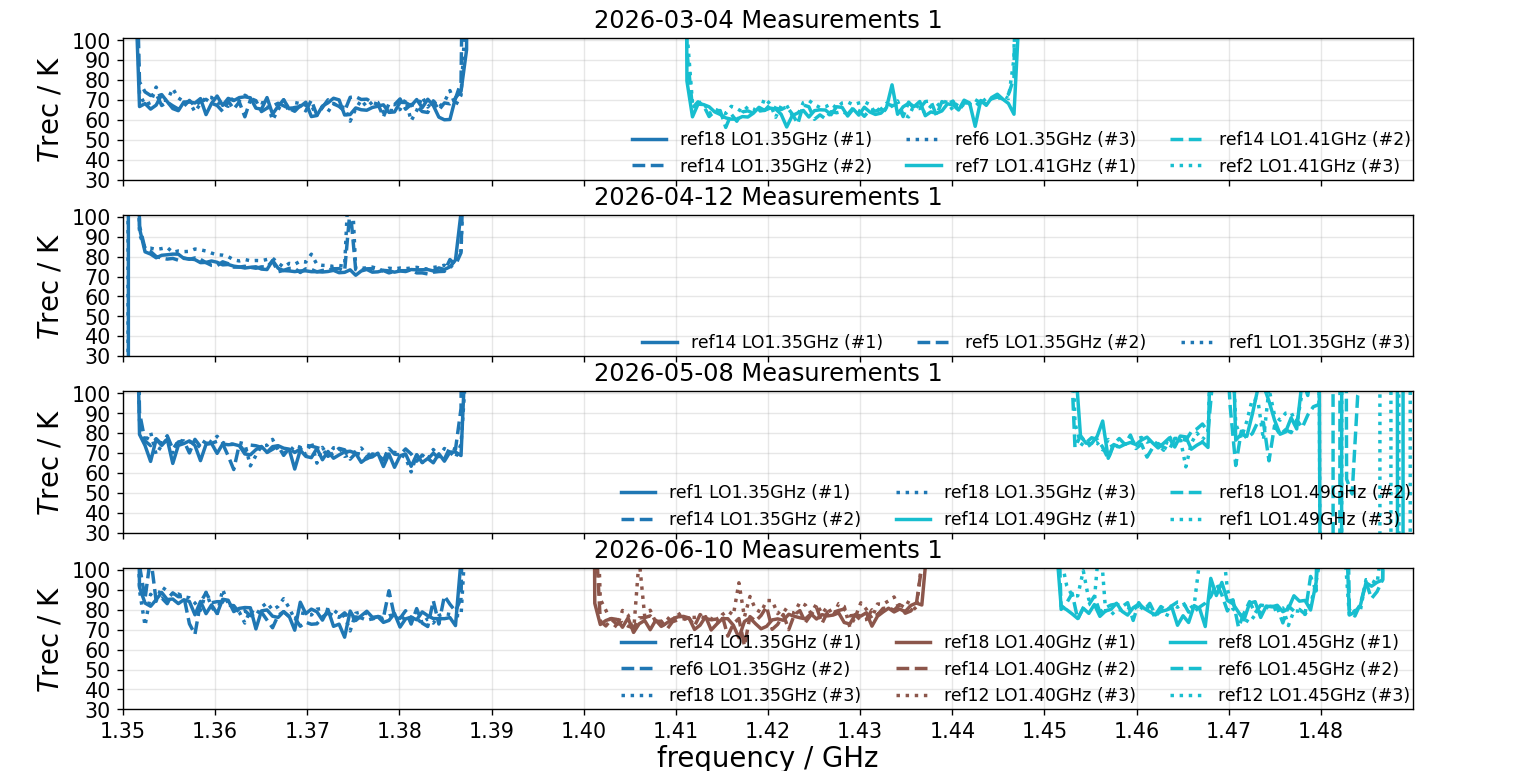}
\caption{The equivalent receiver noise temperature (\(T_{\rm rec}\)) as a function of radio frequency. 
The panels from top to bottom correspond to measurements conducted on 2026 March 04, April 12, May 08, and June 10, respectively. 
The labels in the lower-right corner of each panel indicate the reference antenna index and local oscillator (LO) frequency used 
for array calibration. 
For example, ref18 LO1.35GHz denotes that the measurement adopts reference antenna 18 with an LO frequency of 1.35 GHz.
$(\# 1)$, $(\# 2)$, $(\# 3)$ denotes the top3 averaged \(T_{\rm rec}\) in each measurement.  
Each measurement covers a frequency bandwidth of 38.5 MHz spanning 1350–1490 MHz, with a frequency interval of approximately 0.60 MHz.
The total observation time of each measurement is about 10.5s with time resolution $t_{\rm s}=26.5974$ $\mu$s.  }
\label{fig_4}
\end{figure}

The \(T_{\rm rec}\) results are presented in Fig.\,\ref{fig_4}. 
For the local oscillator (LO) frequency of 1.35 GHz, the lowest (\#1 in Fig.\,\ref{fig_4}) averaged \(T_{\rm rec}\) values are 
\(67.14\pm3.13\ \text{K}\), \(75.36\pm3.88\ \text{K}\), \(70.33\pm3.87\ \text{K}\), and \(78.62\pm4.81\ \text{K}\) 
for the observations on March 04, April 12, May 08, and June 10, respectively. At an LO frequency of 1.41 GHz, 
the measured average \(T_{\rm rec}\) on March 04 is \(65.12\pm3.63\ \text{K}\). 
For the 1.40 GHz LO configuration on June 10, the average \(T_{\rm rec}\) is \(75.15\pm3.28\ \text{K}\). 
In addition, the derived \(T_{\rm rec}\) values at 1.42 GHz are approximately 73 K and 75 K for the March 04 and 
June 10 observations, respectively. For comparison, we also plot the top3 averaged $T_{\rm rec}$ in Fig.\,\ref{fig_4}.
The averaged receiver temperature and standard deviation are listed in Table\,\ref{tab:tsys_top3}. 
Note that when calculating the average values, we discarded anomalous data points with extreme high or low values, 
as well as samples falling outside the operating bandpass. 

As shown in Table\,\ref{tab:tsys_top3}, the standard deviations of measured \(T_\mathrm{rec}\) 
are distributed between approximately 2 K and 8 K for all local oscillator (LO) frequencies. 
However, we can see that in Fig.\,\ref{fig_4} the anomalous data points deviate substantially from the average. 
Part of this discrepancy arises from RFI contamination, while another component can be attributed to aliasing, 
which will be discussed in Section\,\ref{sect:discussion}. The severe signal distortion observed above 1.478 GHz 
originates from the intrinsic bandpass response of element antenna ( 1.05–1.45 GHz), combined with the bandpass 
filter covering 1.37–1.47 GHz. 

An interesting trend can be observed for the LO frequency of 1.35 GHz: the average \(T_\mathrm{rec}\) 
rose from 67 K on March 4 to 75 K on April 12, then decreased to 70 K on May 8, and finally climbed back 
to 79 K on June 10. For the 1.40 GHz LO channel, \(T_\mathrm{rec}\) measured in June is approximately 10 K 
higher than that obtained in March. 
This temporal drift in \(T_\mathrm{rec}\) may be partially attributed to temperature variations of discrete electronic 
components. We measured the physical temperature of the LNA with a thermometer on May 8 and June 10, which were 
37.5 °C and 37.3 °C, respectively. 
When the measurement was conducted on March 4, there were no barriers surrounding the PAF system. 
Afterwards, enclosures made of wood-fiber panels were installed around the PAF to shield the array from 
wind and rain. On April 12, we observed elevated equipment temperature caused by poor air circulation 
inside the enclosure; ventilation improvements including drilled holes were subsequently implemented, 
which yielded a noticeable reduction in the PAF operating temperature within the barriers. The ambient 
temperature reached approximately 16 °C on May 8, considerably lower than the 28 °C ambient temperature 
recorded on June 10. 

According to the manufacturer's datasheet of Mini-Circuits ZX60-P162LN+ 
which was used as the first-stage LNA in the PAF receiver, the noise figure is 0.39\,dB 
at $-45^\circ$C, 0.65\,dB at $25^\circ$C, and 0.84\,dB at $85^\circ$C. 
Using piece-wise linear interpolation, we obtain noise-figure values of 0.61\,dB and 0.69\,dB 
for this LNA at $15^\circ$C and $38^\circ$C, respectively. The corresponding equivalent noise temperatures 
are 43.4\,K and 49.6\,K, yielding an increase of approximately 6\,K. 
The overall receiver noise temperature reaches 67\,K and 79\,K at $15^\circ$C and $38^\circ$C. 
This may suggest that roughly 50\,\% of the receiver-noise variation over the temperature range from 
$15^\circ$C to $38^\circ$C can be attributed to this first-stage LNA.

\begin{table}[htbp]
\centering
\caption{Top-3 references with the minimum average receiver temperature $T_\mathrm{rec}$ under different LO frequencies and measurement dates.}
\label{tab:tsys_top3}
\begin{tabular}{lcccccc}
\hline
\hline
Measurement Date & LO Frequency & Rank & Ref Index & Mean $T_\mathrm{rec}$ (K) & Std. Deviation (K) \\
\hline
2026-03-04 & 1.35 GHz & 1 & 18 & 67.14 & 3.13 \\
2026-03-04 & 1.35 GHz & 2 & 14 & 67.87 & 3.09 \\
2026-03-04 & 1.35 GHz & 3 & 6 & 68.00 & 3.44 \\
2026-03-04 & 1.41 GHz & 1 & 7 & 65.12 & 3.63 \\
2026-03-04 & 1.41 GHz & 2 & 14 & 65.68 & 3.41 \\
2026-03-04 & 1.41 GHz & 3 & 2 & 66.88 & 2.56 \\
\hline
2026-04-12 & 1.35 GHz & 1 & 14 & 75.36 & 3.88 \\
2026-04-12 & 1.35 GHz & 2 & 5 & 75.96 & 5.39 \\
2026-04-12 & 1.35 GHz & 3 & 1 & 79.06 & 8.20 \\
\hline
2026-05-08 & 1.35 GHz & 1 & 1 & 70.33 & 3.87 \\
2026-05-08 & 1.35 GHz & 2 & 14 & 71.89 & 4.06 \\
2026-05-08 & 1.35 GHz & 3 & 18 & 72.02 & 4.25 \\
2026-05-08 & 1.49 GHz & 1 & 14 & 78.50 & 6.60 \\
2026-05-08 & 1.49 GHz & 2 & 18 & 78.54 & 6.93 \\
2026-05-08 & 1.49 GHz & 3 & 1 & 79.01 & 7.21 \\
\hline
2026-06-10 & 1.35 GHz & 1 & 14 & 78.62 & 4.81 \\
2026-06-10 & 1.35 GHz & 2 & 6 & 79.69 & 6.90 \\
2026-06-10 & 1.35 GHz & 3 & 18 & 79.95 & 4.96 \\
2026-06-10 & 1.40 GHz & 1 & 18 & 75.15 & 3.28 \\
2026-06-10 & 1.40 GHz & 2 & 14 & 75.78 & 3.85 \\
2026-06-10 & 1.40 GHz & 3 & 6 & 79.29 & 4.54 \\
2026-06-10 & 1.45 GHz & 1 & 8 & 81.64 & 5.40 \\
2026-06-10 & 1.45 GHz & 2 & 6 & 82.26 & 4.75 \\
2026-06-10 & 1.45 GHz & 3 & 12 & 82.29 & 5.91 \\
\hline
\end{tabular}
\end{table}

\begin{figure}
\hspace*{-1.05cm}
\centering
\includegraphics[width=18cm,clip,angle=0]{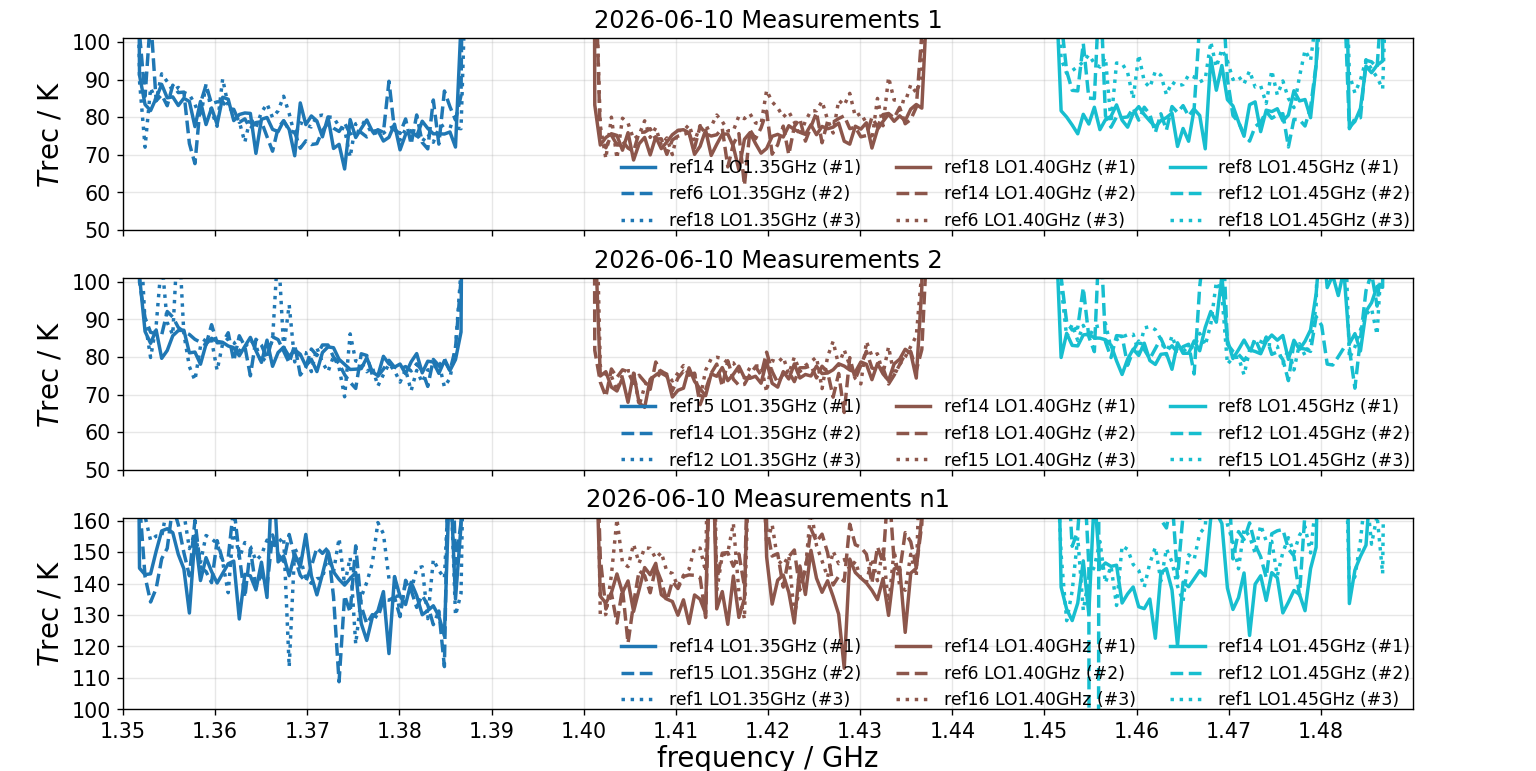}
\caption{The equivalent system noise temperature (\(T_{\rm rec}\)) as a function of radio frequency on June 10. 
The panels from top to bottom correspond to measurements conducted for the cold load with the steel shielding screen (measurements 1), 
backup observation 1-6 mins after measurements 1 (measurements 2), and the cold load without the screen (measurements n1).
The labels in the lower-right corner of each panel indicate the reference antenna index and local oscillator (LO) frequency used 
for array calibration. 
For example, ref14 LO1.35GHz denotes that the measurement adopts reference antenna 14 with an LO frequency of 1.35 GHz.
$(\# 1)$, $(\# 2)$, $(\# 3)$ denotes the top3 averaged \(T_{\rm rec}\) in each measurement.  
Each measurement covers a frequency bandwidth of 38.5 MHz spanning 1350–1490 MHz, with a frequency interval of approximately 0.60 MHz.
The total observation time of each measurement is about 10.5s with time resolution $t_{\rm s}=26.5974$ $\mu$s.  }
\label{fig_5}
\end{figure}

We compare the $T_{\rm rec}$ measurements on June 10 in Fig.\,\ref{fig_5}.  
The panels from top to bottom in this figure correspond to measurements conducted 
for the cold load with the steel shielding screen (measurements 1), 
backup observation 1-6 mins after measurements 1 (measurements 2), and the cold load without the screen 
(measurements n1).
Minor subtle fluctuations of \(T_\mathrm{rec}\) against different reference antenna can be observed 
within several minutes from the figure. Notably, after removing the metal shielding screen, the measured 
\(T_\mathrm{rec}\) significantly increases from approximately 75 K (with the screen) to over 142 K. This 
phenomenon is likely caused by electromagnetic interference coupled through the antenna sidelobes, 
where the noise contributors may be mountainous terrain, vegetation and surrounding buildings nearby. 

The noise received by the antenna array can be decomposed into contributions originating from 
the main beam, sidelobes and backlobes. The noise temperature of the antenna array can be estimated as
$T_A = \eta_{\rm mb} T_{\rm sky} + \eta_{\rm side} T_{\rm side} + \eta_{\rm back} T_{\rm back}$,
where $ \eta_{\rm mb}$, $\eta_{\rm side}$ and $\eta_{\rm back}$ are the main beam, sidelobe beam and backlobe beam efficiencies respectively.  
The beam efficiency $\eta_\star$ ($\star$: mb, side, back) is defined as the ratio of 
the gain-weighted beam solid angle $\Omega_{\star}$ integrated over the corresponding 
angular region to the total array beam solid angle $\Omega_t$, 
i.e. $\eta_{\star} = \frac{\Omega_{\star}}{\Omega_t}$, where $\Omega_{\star}=\int_{\Omega_{\star}} G(\Omega) d\Omega$ 
and $\Omega_t=\int_{4\pi} G(\Omega) d\Omega$, where $G(\Omega)$ is the antenna gain. 
These efficiencies satisfy the normalization condition 
$\eta_{\rm mb} + \eta_{\rm side} + \eta_{\rm back} = 1$.
In the case of stainless-steel screen installed, we can estimate the sidelobe noise temperature 
via $T_{\rm side}=\varepsilon T_{\rm env}+(1-\varepsilon)T_{\rm sky}\approx27$ K, where $\varepsilon$ is the L-band 
emissivity of stainless steel ($\varepsilon \approx 0.07$), $T_{\rm env}=301$ K is the environmental temperature on June 10, 
and $(1-\varepsilon)T_{\rm sky}\approx6$ K represents the cold-sky radiation reflected by the inner surface of the steel screen. 
We here take $\eta_{\rm back}\approx0.03$ \citep{Wu2013} and $T_{\rm back}\approx301$ K. 
When the steel screen was removed, the emissivity becomes $\varepsilon \approx1.0$, and $T_{\rm side}\approx301$ K due to 
the surrounding mountains. 
If we assume that the increase of sidelobe-induced noise from the screened and unscreened case results in \(67\ \text{K}\) 
rise in the receiver noise, this would suggest a beam efficiency of \(0.76\), which is lower than the 
mean value ($\sim$0.88) of beam efficiency in \citet{Chippendale2014} by 15.8\%. 
In the next work, we will pay attention to the radiation-pattern measurement to understand the beam efficiency and antenna noise budget.

\subsection{Influences of beamforming weights}
\label{subsec:beamformers}

\begin{figure}
\hspace*{-1.05cm}
\centering
\includegraphics[width=18cm,clip,angle=0]{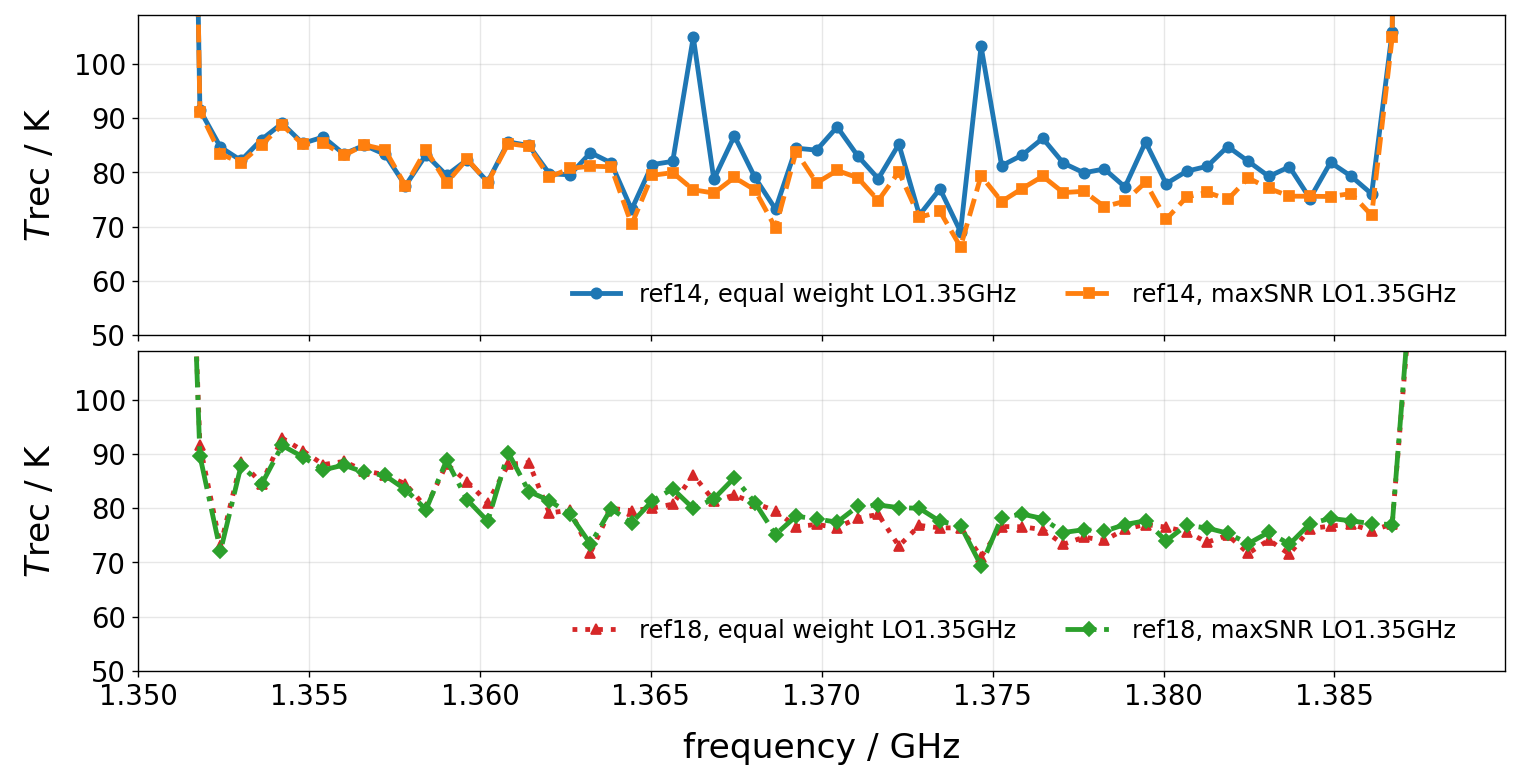}
\caption{The equivalent receiver noise temperature (\(T_{\rm rec}\)) as a function of radio frequency on June 10 under LO 1.35GHz. 
The top panel is for the reference antenna 14 with equal weight and maxSNR weight, while the bottom panel is for the reference 
antenna 18.  
Each measurement covers a frequency bandwidth of 38.5 MHz, with a frequency interval of approximately 0.60 MHz.
The total observation time of each measurement is about 10.5s with time resolution $t_{\rm s}=26.5974$ $\mu$s. 
The solid circles, squares, triangles and diamonds denote the measured data points at the corresponding frequencies. }
\label{fig_6}
\end{figure}

We compare the influences of different reference antennas and beamforming weights on receiver noise in Fig.\,\ref{fig_6}.
We find that compared with equal-weight beamforming, the maxSNR beamforming combined with the noise-whitening method 
can suppress receiver noise and yields estimates with lower variance and reduced bias. 
This is in line with previous studies (e.g. \citet{Landon2010}. 
Furthermore, our results indicate that the performance gain of the maxSNR beamforming method may be diminished 
when the antenna-calibrated measurement data inherently exhibits low variance.
Future work will include further comparisons of additional beamforming strategies 
and their impacts on \(T_{\rm sys}\) measurements.

\subsection{Mutual coupling}
\label{subsec:tmc}

\begin{figure}
\hspace*{-1.1cm}
\centering
\includegraphics[width=18cm,clip,angle=0]{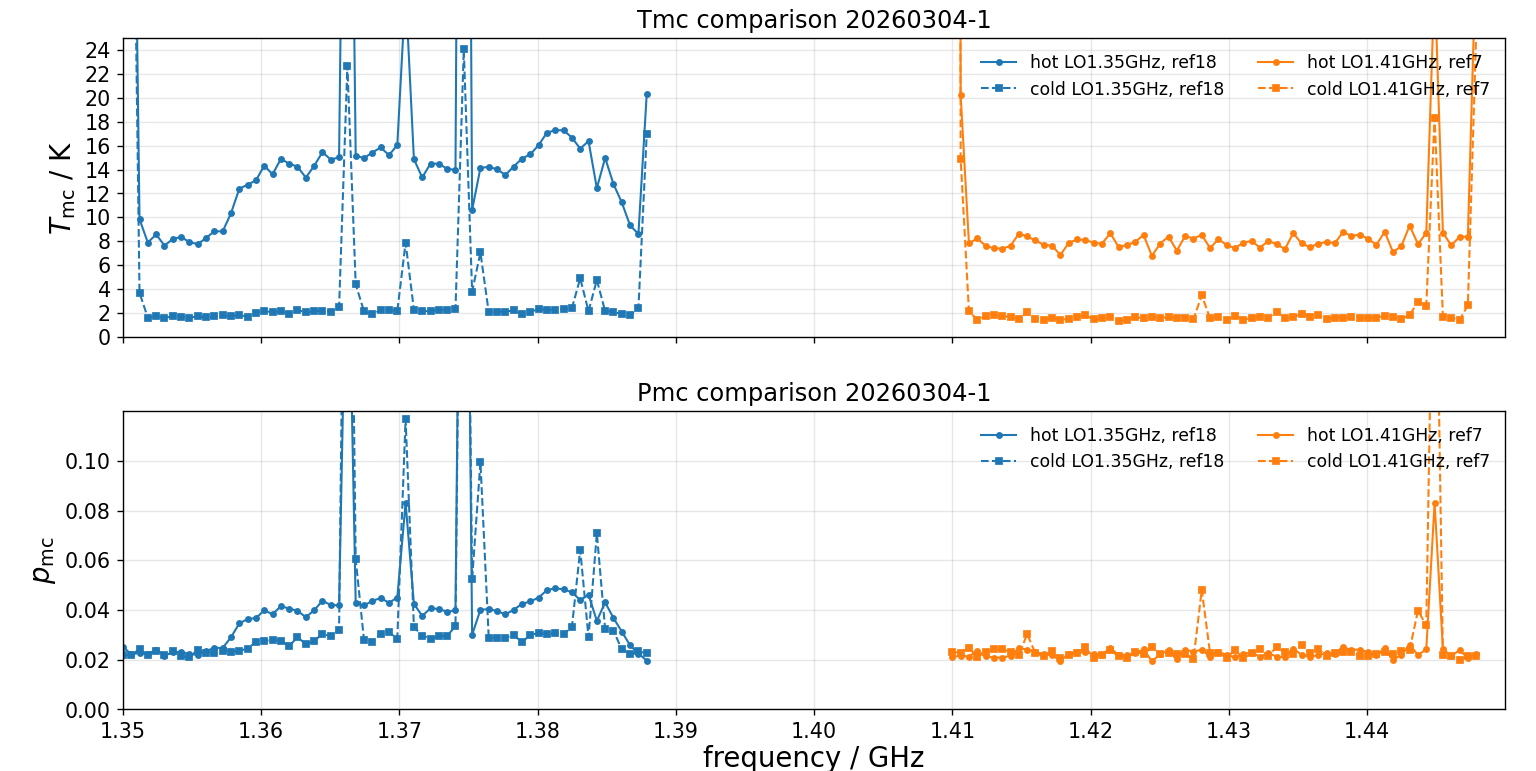}
\caption{The mutual coupling noise (\(T_{\rm mc,\ast}\)) (top panel) and the 
coupling proportion $p_{\rm mc,\ast}$ (bottom panel) as a function of radio frequency 
on 2026 March 04. 
Labels in this figure, namely LO* and ref* denote the LO frequency and reference antenna used in the measurements, 
e.g., LO1.35GHz and ref18 means the LO frequency is 1.35GHz, and the number of reference antenna is 18.   
Each measurement covers a frequency bandwidth of 38.5 MHz, with a frequency interval of approximately 0.60 MHz.
The solid and dashed lines are for the \( \ast =\) hot and cold respectively. 
The solid circles, squares, triangles and diamonds denote the measured data points at the corresponding frequencies. 
The total observation time of each measurement is about 10.5s with time resolution $t_{\rm s}=26.5974$ $\mu$s. 
}
\label{fig_7}
\end{figure}

\begin{figure}
\hspace*{-1.1cm}
\centering
\includegraphics[width=18cm,clip,angle=0]{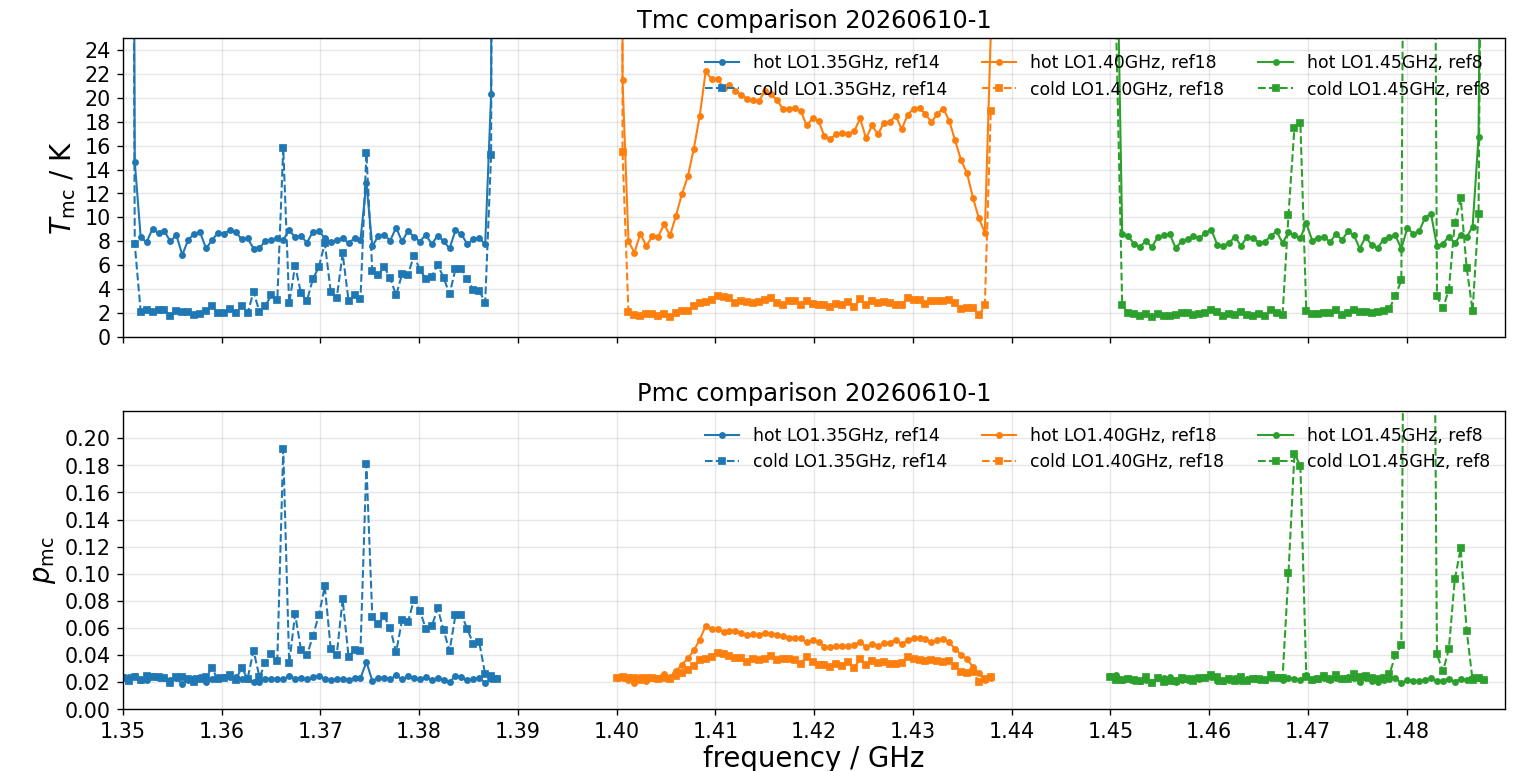}
\caption{Same with Fig.\,\ref{fig_7}, but the date of measurements took place was 2026 June 10. 
}
\label{fig_8}
\end{figure}

We plot mutual coupling temperature \(T_{\rm mc}\) and coupling proportion 
\( p_{\rm mc}\) (sum of the off-diagonal entries relative to the 
total sum of all entries within the cold-load covariance matrix \(\mathbf{R}_\mathrm{cold}\) and 
hot-load covariance matrix \(\mathbf{R}_\mathrm{hot}\) ) in Fig.\,\ref{fig_7} (2026 March 04) 
and \ref{fig_8} (2026 June 10). 
 
Figure \ref{fig_7} illustrates \(T_{\rm mc}\) and \(p_{\rm mc}\) across two LO frequencies 
spanning 1.35–1.45 GHz, with paired hot (solid lines, circle markers) and cold (dashed lines, square markers) 
measurements for reference antenna ref18 (LO1.35GHz: LO 1.35GHz) and ref7(LO1.41GHz: LO 1.41GHz).
The measurements on Fig.\,\ref{fig_8} uses three LO frequencies 1.35GHz, 1.40GHz, and 1.45GHz, and 
corresponding references antennas are ref14, ref 18 and ref8. 

In Fig.\,\ref{fig_7}, when LO frequency is 1.35GHz, hot \(T_{\rm mc}\) starts at $\sim$8 K, 
gradually climbs to 14–18 K between 1.37–1.38 GHz, and features tall, narrow resonant spikes above 22 K. 
Cold \(T_{\rm mc}\) remains close to 0–2 K with smaller matching resonant peaks. 
The \(p_{\rm mc}\) curve reflects this behavior: a slowly rising baseline around 0.02–0.05 for hot calibration, 
plus sharp resonant peaks exceeding 0.10, while cold signals stay weak with narrow spikes.
For LO frequency 1.41GHz, hot \(T_{\rm mc}\) forms a flat, stable baseline near 8 K with only 
a few isolated narrow resonances, while cold \(T_{\rm mc}\) is constrained to $\sim$0–2 K. 
The \(p_{\rm mc}\) baseline is uniformly low at $\sim$0.02 for both hot and cold, with only minor narrow resonant 
spikes at select frequency points.
Similar to the LO1.35GHz dataset, hot calibration produces systematically stronger mutual coupling signatures than cold 
calibration across both bands, and resonant coupling appears as discrete narrow peaks superimposed on smooth, slowly 
varying baselines.

In Fig.\,\ref{fig_8}, for LO1.35GHz, hot \(T_{\rm mc}\) maintains a stable baseline around 8–10 K, 
with sharp narrow resonant peaks up to 16 K; 
cold \(T_{\rm mc}\) stays much lower at roughly 0–4 K. 
The corresponding \(p_{\rm mc}\) follows the same trend: the hot curve hovers near 0.02 with resonant peaks 
reaching 0.20, while the cold signal exhibits weaker but consistent resonant features. 
The LO1.40GHz setup delivers distinctly elevated coupling signals. 
Its hot \(T_{\rm mc}\) rises to a broad plateau of 16–22 K, far higher than the other two bands, 
with a smooth, wide hump rather than narrow discrete resonances. 
Cold \(T_{\rm mc}\) remains suppressed at ~2–4 K. \(p_{\rm mc}\) shows a broad, mild peak around 
0.06 for hot calibration and 0.04 for cold, without sharp narrow spikes. 
When using LO1.45GHz, the overall coupling returns to levels comparable to LO1.35GHz. 

Across all LO frequency setup, hot calibration consistently yields significantly larger 
\(T_{\rm mc}\) and \(p_{\rm mc}\) values than cold calibration, indicating that hot-load 
measurement amplifies observable mutual coupling effects. However, we find that for the 
measurement of LO1.35GHz on June 10, at many frequency bins, cold \(p_{\rm mc}\) values 
are larger than hot \(p_{\rm mc}\). This phenomenon could be caused by RFI, and further 
measurements are necessary to understand such phenomenon. 

The measurements on both dates demonstrate that: 
1. hot-load measurements produce larger 
\(T_{\rm mc}\) and \(p_{\rm mc}\) than cold-load measurements. 
This validates that hot calibration provides higher signal contrast to characterize 
receiver mutual coupling; 
2. We can see that a smooth, frequency-dependent baseline coupling modulated by sharp, narrow resonant spikes 
probably originating from standing waves or array element crosstalk; 
3. low-frequency bands (1.35–1.39 GHz) consistently show stronger frequency evolution of the baseline 
coupling than higher-frequency bands (1.40+ GHz) in the measurements.

The consistent gap between hot and cold traces across both datasets confirms that 
the measured \(T_{\rm mc}\) and \(p_{\rm mc}\) can reflect the receiver’s intrinsic mutual 
coupling. The dramatic variation among measurements with LO bands and different reference channels 
demonstrates that mutual coupling is highly sensitive to reference hardware and front-end assembly. 
The hot and cold measurements capture distinct hardware configurations, enabling us to disentangle 
the contributions of mutual coupling to the overall performance of the PAF receiver system.

\subsection{Equivalent system noise temperature}
\label{subsec:tsys}

\begin{figure}
\hspace*{-1.05cm}
\centering
\includegraphics[width=18cm,clip,angle=0]{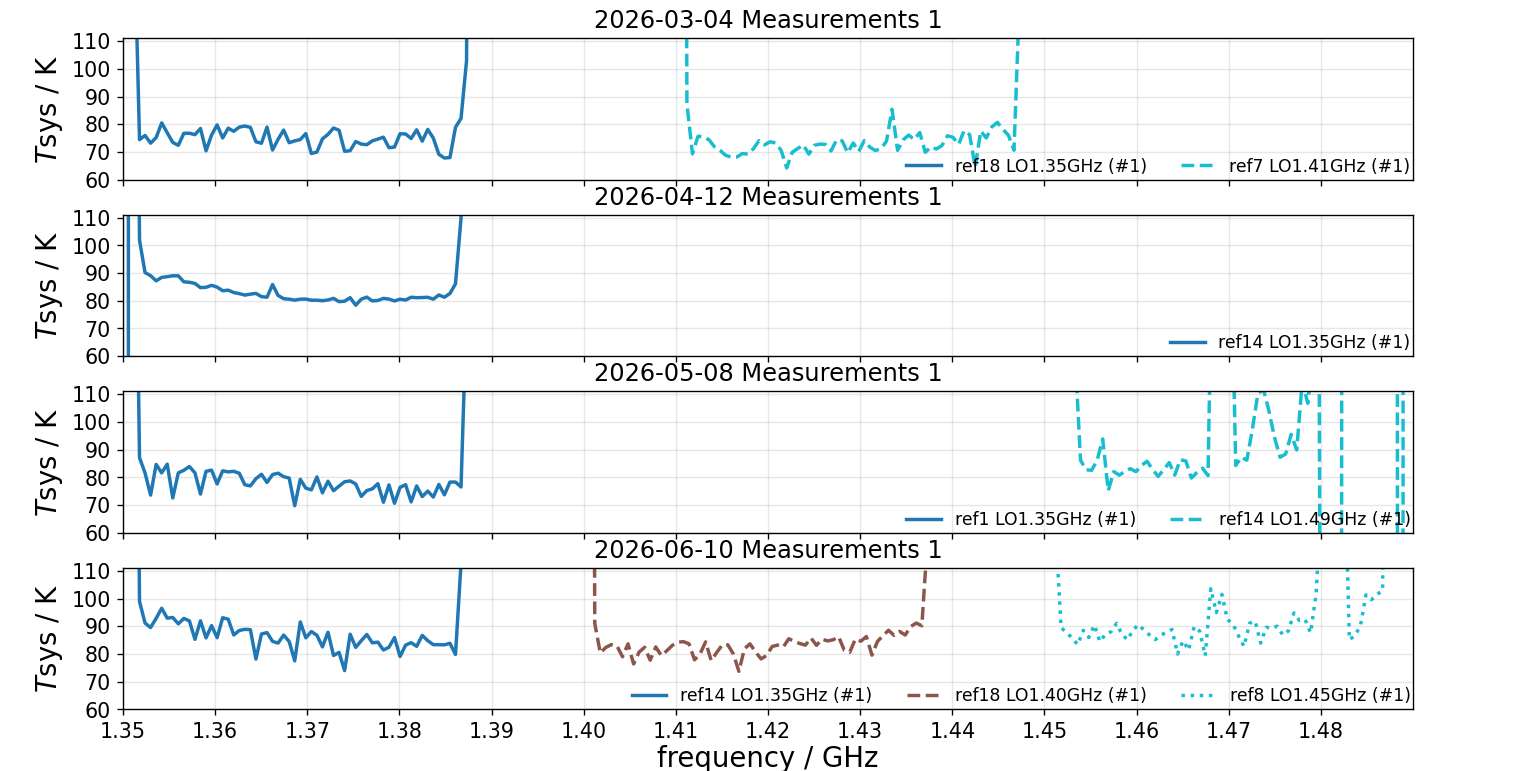}
\caption{The equivalent system noise temperature ($T_{\rm sys}$) is plotted as a function of radio frequency. 
The panels from top to bottom correspond to measurements conducted on 2026 March 04, April 12, May 08, and June 10, respectively. 
The labels in the lower-right corner of each panel indicate the reference antenna index and local oscillator (LO) frequency 
used for array calibration. For example, ref18 LO1.35GHz denotes that the measurement adopts reference antenna 18 with an LO frequency 
of 1.35 GHz. Each measurement covers a frequency bandwidth of 38.5 MHz spanning 1350–1490 MHz, with a frequency interval of approximately 0.60 MHz. 
The total observation time of each measurement is about 10.5s with time interval $t_{\rm s}=26.5974$ $\mu$s.}
\label{fig_3}
\end{figure}

The measured system noise temperature \(T_\mathrm{sys}\) is shown in Fig.\,\ref{fig_3}.
For the local oscillator (LO) frequency of 1.35 GHz, the averaged \(T_{\rm sys}\) values are \(74.87\pm3.13\ \text{K}\), 
\(83.09\pm3.88\ \text{K}\), \(78.06\pm3.87\ \text{K}\), and \(86.35\pm4.81\ \text{K}\) for the observations on March 04, 
April 12, May 08, and June 10, respectively. 
At an LO frequency of 1.41 GHz, the measured average \(T_{\rm sys}\) on March 04 is \(72.85\pm3.63\ \text{K}\). 
For the 1.40 GHz LO configuration on June 10, the average \(T_{\rm sys}\) is \(82.88\pm3.28\ \text{K}\). 
In addition, the derived \(T_{\rm sys}\) values at 1.42 GHz are approximately 81 K and 83 K for the March 04 and June 10 observations, respectively. 

The averaged equivalent system noise temperature and the main components over the frequency 
band 1350-1390MHz for the measurements on 2026 March 04 and June 10 are shown in Table \ref{tab_Tsys}.

\begin{table*}
\caption{Averaged equivalent system noise temperature and the main components over the frequency 
band 1350-1390MHz for the measurements on 2026 March 04 and June 10 in our observations, adopting 
ambient temperature of 288K and maxSNR beamformer. }
\label{tab_Tsys}
\begin{center}
\begin{tabular}{lccccccc}
\hline
                   &  2026 March 04 LO1.35GHz ref18 &  2026 June 10 LO1.35GHz ref14  \\
 \hline
 $\tilde{T}_{\rm rec}$     &   \(55\pm3 \text{K}\) (measured)       & \(64\pm5\ \text{K}\)  (measured)  &\\
 $T_{\rm mc, cold}$       &    \( 2\pm1  \text{K} \) (measured)  &    \( 4\pm2  \text{K} \) (measured) &     \\
 $T_{\rm sky}$           &   5.73K (estimated)   &    5.73K (estimated)     &\\
  $T_{\rm loss}$        & 10K (estimated)     &   10K (estimated)     & \\
   $T_{\rm sp}$        &  2K (estimated)             & 2K (estimated)             & \\
    \hline
      $T_{\rm sys}$     &  75$\pm$3K   &     \(86\pm5\ \text{K}\)           & \\
\hline
\end{tabular}
\end{center}
\end{table*}

\section{Discussion}
\label{sect:discussion}

Aliasing affects the measurements with LO frequencies of 1.40–1.45 GHz. 
Taking LO = 1.40 GHz as an example, signals from 1.36–1.40 GHz alias into the plotted 1.40–1.44 GHz band. 
Owing to such spectral overlap, the corresponding noise data can only act as a reference for the overall 
system noise performance across the bandwidth of 1.36–1.45 GHz. 
In our case, the receiver output power $P_{\mathrm{out},*}$ under aliased conditions can be written as
$P_{\mathrm{out},*}=k_B G\Delta f_1 \left(T_{*}+T_\mathrm{rec}\right) + k_B G\Delta f_2 \left(T_{*}+T_\mathrm{rec}\right)$, 
where $k_B$ is the Boltzmann constant, $G$ denotes the receiver gain, and $\Delta f_1$, $\Delta f_2$ are the 
bandwidths corresponding to the negative-frequency and positive-frequency components, respectively. 
We find that the Y-factor $Y=P_{\mathrm{out,hot}}/P_{\mathrm{out,cold}}=(T_\mathrm{hot}+T_\mathrm{rec})/(T_\mathrm{sky}+T_\mathrm{rec})$ 
remains unchanged. The measured results are consistent with this simplified theoretical analysis. 
Note that the presence of strong RFI may influence the Y-factor measurement. 
We will upgrade and reconfigure hardware in subsequent measurement campaigns to suppress this spectral folding effect. 

We can alternatively calculate the receiver noise temperature due to the first stage LNAs and subsequent 
circuits by the cascaded noise factor formula

\begin{equation}
\label{eq:ftotal}
F_{\rm total}=F_1+\frac{F_2-1}{G_1}+\frac{F_3-1}{G_1G_2}+\frac{F_4-1}{G_1G_2G_3}+...
\end{equation}
where $F_1$, $F_2$, $F_3$ denotes the noise factor of the first stage amplifier, second stage amplifier, 
third stage amplifier, and $G_1$, $G_2$, $G_3$ the gain of each stage amplifier respectively.
From the formula, we see that if the noise factor is sufficiently small or the gain is large, 
the noise temperature of a cascaded system is mainly dominated by the first stage LNA.

We used the ZX60-P162LN+ manufactured by Mini-circuits as the first stage LNA. It has a noise figure of 
0.6 dB in room temperature and gain of $\sim$20 dB at about 1.4GHz. The measured total equivalent receiver 
noise figure using a noise source in front of the first level LNA is about 0.67 dB , which lead to a noise 
temperature of about 48 K. The gain, noise figure and noise temperature contribution of each amplifier (typical) 
are listed in table\,\ref{tab:estimatedtrec}. 
This result is approximately consistent with our noise temperature measurements.

\begin{table*}
\caption{The parameters of main amplifiers used in the PAF and their contributes to total NF and $T_{\rm rec}$. }
\begin{center}
\begin{tabular}{ccccc}
\hline
\hline
name & gain & noise figure & contribute to NF & contribute to $T_{\rm rec}$\\
\hline
ZX60-P162LN+& 20dB& 0.6dB& 0.6dB& 43K\\
ZRL-2150+& 25dB& 1.5dB& 0.053dB& 4K\\
ZFL-2500VHX+& 20dB& 5.5dB& 0.0013dB& 0.15K\\
\hline
\hline
\label{tab:estimatedtrec}
\end{tabular}
\end{center}
\end{table*}

Simulations and S-parameter measurements of the six inner-ring antenna elements 
in the PAF \citep{Wu2013} show that the maximum transmission coefficient \(S_{1,2}\) 
between the central element and an adjacent inner-ring element reaches -18 dB and -24 dB. 
Given an ambient temperature of 288 K, the corresponding mutual coupling noise temperatures 
are estimated to be approximately 4.6 K and 1.2 K respectively. 
S-parameter simulations and measurements for the central element and 12 outer-ring elements 
show \(-40 \text{dB} < S_{1,9} < -30 \text{dB} \), yielding a mutual coupling noise contribution 
in the range of $\sim$0.03-0.3 K. The overall mutual coupling noise temperature of the full array 
ranges approximately from 7.6 K to 10.7 K.
Our measurements agree well with S-parameter data, and further indicate that mutual coupling's 
contribution to system noise temperature depends on ambient radio sky temperature and RFI levels. 

\citet{Chippendale2014} presented system noise temperature measurements for phased-array antennas 
operating in receiving band from 0.7 GHz to 1.8 GHz. 
Similar to our experiments, they also used a large microwave absorber at ambient temperature and the
unobstructed radio sky as the hot and cold load. 
The main differences between their work and ours are: 1) they used a broadband reference antenna as 
noise source centred over and pointed at the array under test, which can make noise measurements with 
reproducible, well-defined beams directed at the zenith and thereby at the centre of the absorber target; 
2) our multi-band, multi-epoch measurements utilise deployable shielding screens to isolate sidelobe spillover 
noise; 3) we propose a covariance-based method to extract mutual coupling temperature \(T_{\rm mc}\).  
Validated against S-parameter simulations and tests, our work further identifies the dependence of coupling noise 
on radio frequency interference (RFI) and radio sky temperature, a finding unreported in the CSIRO research. 
Additionally, our pipeline incorporates iterative channel calibration and long-term multi-epoch experiments over months, 
whereas the CSIRO workflow focuses on single experiment. 

\section{Conclusion}
\label{sect:conclusion}

In this paper, we present comprehensive system noise temperature measurements of a 19-element hexagonal room-temperature PAF 
via the modified Y-factor method across multiple frequency bands and multi-month observation campaigns. 
We developed a complete Python-based data reduction pipeline containing DC offset removal, iterative inter-channel mutual 
calibration and maxSNR beamforming to do the data reduction.
A detachable stainless steel shielding screen was deployed to quantitatively isolate sidelobe spillover noise originating 
from terrain and surrounding environments. Comparative measurements demonstrate that unshielded sidelobe pickup elevates 
the receiver noise temperature by more than 60 K, clearly verifying the contamination of spillover radiation to noise 
characterization. 

Based on the covariance matrix off-diagonal terms extracted from hot/cold calibration datasets, a quantitative decomposition 
scheme for mutual coupling noise temperature \(T_{\rm mc}\) was proposed as shown in Sec.\,\ref{sec:myfactor}. 
The baseline of the measured mutual coupling noise temperature ranges from approximately 2 K to 8 K, and can 
reach up to 20 K under certain observational conditions.
Cross-validation with element-to-element S-parameter 
results shows that inner adjacent elements with \(S_{21} = -18\sim-24\ \text{dB}\) generate coupling noise of 1.2–4.6 K, 
while distant outer-ring elements (\(S_{21} < -30\ \text{dB}\)) only contribute approximately 0.03–0.3 K. 
The integrated mutual coupling noise of the full array falls within 7.6–10.7 K, which also matches well with electromagnetic 
simulation predictions. 

Consistent offsets of \(T_{\rm mc}\) between hot and cold calibration states confirm that hot-load observations provide higher signal 
levels of inter-element crosstalk by contrast with cold-load. 
We further reveal that mutual coupling noise is jointly induced by the thermal radiation associated with the ambient 
radio-sky temperature and non-thermal radiation most likely originating from RFI, an effect unreported in previous PAF 
noise measurement studies including \citet{Chippendale2014}.
Long-term monitoring over March–June 2026 captures obvious temporal drift of receiver noise, which may be mainly attributed to the 
temperature fluctuation of front-end LNAs and heat dissipation conditions inside the protective enclosure. 

The cascaded noise factor calculation indicates that the first-stage LNA dominates the receiver noise budget, consistent with our 
measured \(T_{\rm rec}\) ranging from 65 K to 82 K under different LO configurations. Combined with sky background, spillover, loss 
and mutual coupling contributions, the overall beam-equivalent system noise temperature of the PAF is measured to be $\sim$73–89 K within 
the 1.37–1.47 GHz operating band. The measurement framework, channel calibration workflow and mutual coupling decomposition method 
established in this work offer a practical reference for noise performance evaluation of room-temperature PAF prototypes for ground-based 
radio telescopes. 
In future work, we will compute adaptive beamforming weights from long-term covariance matrix measurements and 
measure the radiation pattern of the PAF receiver to jointly characterize the temporal variation of system noise.

\normalem
\begin{acknowledgements}

We thank the reviewer for the helpful comments. 
This work was supported by the Science and Technology Innovation Program
of Hunan Province (No. 2024JC0001 No. 2025RC4002), and the Investigation 
of Technological Infrastructure Resources (No. 2023FY101101). 
This work was also supported by the CAS Project for Young Scientists 
in Basic Research, Grant No. YSBR-063 
and the National Natural Science Foundation of China (Grant No. 11073027, 11673031, 12373104). 
We are grateful to Mr. Shengyi Zhou (People's Government, Longshan County) and Mr. Jie Zhu 
(People's Government of Guitang Town) for their strong local supports. 

\end{acknowledgements}

\bibliographystyle{raa}
\bibliography{bibtex}

\end{document}